\documentclass[fleqn,twoside]{article}
\usepackage{espcrc2}

\usepackage{graphicx}
\usepackage{epsfig}
\usepackage[figuresright]{rotating}

\newcommand{\be}{\begin{equation}}                                              
\newcommand{\ee}{\end{equation}}
\newcommand{\half}{\frac{1}{2}}
\newcommand{\LSB}{\raisebox{-0.3ex}{\mbox{\LARGE$\left[\right.$}}}
\newcommand{\RSB}{\raisebox{-0.3ex}{\mbox{\LARGE$\left.\right]$}}}

\newcommand{\AmS}{{\protect\the\textfont2
  A\kern-.1667em\lower.5ex\hbox{M}\kern-.125emS}}

\title{Unquenched domain wall quarks with TSMB}

\author{qq+q Collaboration \\[0.5em]
        F. Farchioni\address{Institut f\"ur Theoretische Physik,
        Universit\"at M\"unster, 
        Wilhelm-Klemm-Str. 9, D-48149 M\"unster, Germany},
        C. Gebert\address[DESY]{Deutsches Elektronen-Synchrotron, DESY,
        Notkestr. 85, D-22603 Hamburg, Germany},
        I. Montvay\addressmark[DESY]\thanks{Poster presented by
        I. Montvay.},
        L. Scorzato\addressmark[DESY]\\[0.5em] }

\begin{document}

\begin{abstract}
 The numerical simulation of domain wall quarks with the two-step
 multi-boson (TSMB) algorithm is investigated.
 Tests are performed on a $8^3 \cdot 4$ lattice with $N_f=2$ quark
 flavours.
\vspace{1pc}
\end{abstract}

\maketitle

\section{INTRODUCTION}
 Lattice actions with improved chiral properties offer the perspective
 of QCD simulations with a better control of chiral symmetry at
 non-zero lattice spacings.
 A prototype is the domain wall fermion action
 \cite{KAPLAN,NARANEU,SHAMIR} defined on a five dimensional hypercubic
 lattice.
 Following Shamir's formulation \cite{SHAMIR}, the light chiral fermion
 modes are located on two boundaries of the fifth dimension.
 The price of the chiral symmetry at non-zero lattice spacing is the
 extra dimension enlarging the number of degrees of freedom.
 From a technical point of view this means that the extensions of the
 fermion matrix are larger and therefore the numerical simulations
 are slower.

 It is an interesting question how much the computation speed decreases
 compared to, say, Wilson fermions.
 Since the good chiral properties of domain wall fermions develop
 only sufficiently close to the continuum limit, the performance
 studies have to be finally performed on large lattices.
 A comparison can only be conclusive if the autocorrelations of
 important physical quantities are also determined -- a difficult
 task requiring a substantial amount of computer time.

 A first step towards the effective simulation of domain wall quarks
 is to identify possible simulation algorithms which have the
 potential of being applicable in simulations with small quark masses
 and in large physical volumes.
 The two-step multi-boson algorithm (TSMB) \cite{TSMB}-\cite{SYMREV}
 has been succesfully tested in this respect in case of the
 Wilson quark action \cite{LIGHTQ,LIGHTB}.
 The application of TSMB for domain wall quarks has been recently
 considered in \cite{DOMAINWALL}.
 Here we report on some further tests along these lines.

\section{ACTION AND ALGORITHM}
 The lattice action for domain wall quarks can be introduced as
\be \label{eq01}
S = S_G[U] + S_F[\overline{\Psi},\Psi,U] 
+ S_{PV}[\Phi^\dagger,\Phi,U] \ .
\ee
 Here $S_G$ denotes the pure gauge-field part depending on the gauge
 field $U$, $S_F$ is the fermionic part with the Grassmannian
 quark fields $\Psi,\overline{\Psi}$ and $S_{PV}$ depends on the bosonic
 Pauli-Villars field $\Phi$ which subtracts the heavy fermion modes --
 as introduced in \cite{VRANAS}.

 The fermion action $S_F$ is defined by
\be \label{eq02}
S_F = \sum_{x,s;\,x^\prime,s^\prime} \overline{\Psi}(x^\prime,s^\prime)
D_F(x^\prime,s^\prime;\,x,s) \Psi(x,s) \ .
\ee
 The four-dimensional space-time coordinates are denoted by $x,x^\prime$
 and the fifth coordinates are $s,s^\prime$.
 The domain wall fermion matrix $D_F$ is constructed from the standard
 four-dimensional Wilson fermion matrix
\begin{eqnarray} &&
D(x^\prime,x) =
\delta_{x^\prime x}\; (4-am_0) 
\nonumber
\\ &&
- \half \sum_{\mu=1}^4 \LSB
\delta_{x^\prime,x+\hat{\mu}}(1+\gamma_\mu) U_{x\mu}
\nonumber
\\[-0.4em] && \label{eq03}
+\; \delta_{x^\prime+\hat{\mu},x}(1-\gamma_\mu) U^\dagger_{x^\prime\mu}
\RSB \ .
\end{eqnarray}
 The notations are standard: $a$ is the (four-dimensional) lattice
 spacing and $\hat{\mu}$ denotes the unit vector in direction $\mu$.
 The bare mass $-m_0$ is chosen to be negative and should be tuned
 properly for producing the light boundary fermion state.
 The non-zero matrix elements of the domain wall fermion matrix are:
\vspace*{-0.3em}
\begin{eqnarray} &&
(D_F)_{s,s} = \sigma+D \ ,\;
(D_F)_{s,s+1} = -\sigma P_L \ ,
\nonumber
\\ &&
(D_F)_{s+1,s} = -\sigma P_R \ ,\;
(D_F)_{1,N_s} = am_f P_R \ ,
\nonumber
\\ &&
(D_F)_{N_s,1} = am_f P_L \ .
\label{eq04}
\end{eqnarray}
 Here $m_f$ denotes the bare fermion mass of the light boundary fermion,
 $N_s$ is the lattice extension in the fifth dimension,
 $P_R = \half(1+\gamma_5),\; P_L = \half(1-\gamma_5)$ are chiral
 projectors and $\sigma \equiv a/a_s$ determines the lattice spacing in
 the fifth dimension $a_s$ relative to $a$.

 The TSMB algorithm is based on the hermitean domain wall fermion matrix
\be \label{eq05}
\tilde{D}_F \equiv \gamma_5 R_5\, D_F
\ee
 where the reflection in the fifth dimension is defined as
 $(R_5)_{s^\prime,s} \equiv \delta_{s^\prime,N_s+1-s}\;
 (1 \leq s,s^\prime \leq N_s)$.
 The quark determinant with $N_f$ flavours is realized in TSMB by
\vspace*{-0.3em}
\be \label{eq06}
\left|\det(\tilde{D}_F)\right|^{N_f} \;\simeq\;
\frac{1}{\det P^{(1)}_{n_1}(\tilde{D}_F^2)\;
\det P^{(2)}_{n_2}(\tilde{D}_F^2)} \ .
\ee
 The polynomials $P^{(1)}$ and $P^{(2)}$ satisfy
\be \label{eq07}
\lim_{n_2 \to \infty} P^{(1)}_{n_1}(x)P^{(2)}_{n_2}(x) =
x^{-N_f/2} \ ,\;
x \in [\epsilon,\lambda] \ .
\ee
 The interval $[\epsilon,\lambda]$ covers the spectrum of
 $\tilde{D}_F^2$ on a typical gauge configuration.
 The first polynomial $P^{(1)}$ of order $n_1$ is a crude approximation
 and is realized by the multi-boson representation \cite{MULTIB}.
 The second polynomial $P^{(2)}$ of order $n_2 \gg n_1$ is a better
 approximation.
 It is taken into account in the updates by a global noisy Metropolis
 correction step.
 Since for fixed $n_2$ (and outside the interval $[\epsilon,\lambda]$)
 the approximation $P^{(1)}P^{(2)}$ is not exact, a final correction
 step is performed by reweighting the gauge configurations which are
 considered for the evaluation of the expectation values.
 (For mode details on TSMB see \cite{TSMB}-\cite{SYMREV}.)

\section{SIMULATION TESTS}
 The implementation of TSMB for domain wall quarks is straightforward.
 Since domain wall fermions have an extra index labeling the fifth
 coordinate, a potential problem is the storage of the $n_1$ multi-boson
 fields in computer memory.
 This problem can be easily solved because (for domain wall fermions
 and in general) one can organize the gauge field update in such a way
 that the dependence on the multi-boson fields is collected in a few
 auxiliary $3 \otimes 3$ matrix fields which can be easily stored in
 memory.
 The multi-boson fields themselves can be kept on disk and have to be
 read before and written back to disk after a complete boson field
 update.
 The duration of the input-output is negligible compared to the time
 of the update.
 Organized in this way, TSMB has a rather low storage requirement.

\begin{figure}[htb]
\vspace{-5.5em}
\begin{flushleft}
\epsfig{file=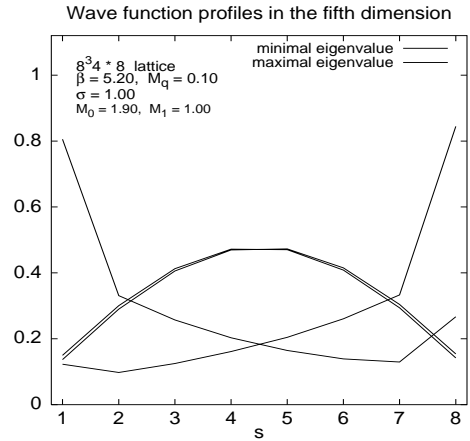,
        bbllx=0pt,bblly=0pt,bburx=510pt,bbury=680pt,
        width=6cm,height=9cm,angle=-90}
\end{flushleft}
\vspace{-1.5em}
\caption{\label{fig:profile}
 Wave function profiles for eigenstates of the hermitean domain wall
 fermion matrix.}
\vspace{-2.0em}
\end{figure}
 We performed test runs for two degenerate quark flavours $N_f=2$
 on $8^3 \cdot 4$ lattices in the vicinity of the $N_t=4$ thermodynamic
 crossover.
 The parameter sets have been chosen from the points in parameter space
 which were investigated in \cite{NT4}.
 Typical parameters were:
 $M_0 \equiv am_0=1.9,\; M_q \equiv am_f=0.1,\;\sigma=1.0,\,0.5$ and
 $5.20 \leq \beta \leq 5.45$.
 (The Pauli-Villars mass was $M_1 = 1$.)
 After equilibrating the gauge configurations several features of the
 domain wall quarks were investigated.
 For typical wave function profiles in the fifth dimension with
 $N_s=8$ see figure \ref{fig:profile}.
 The curves peaking at the walls correspond to eigenvalues with smallest
 absolute value, those concentrated in the middle to eigenvalues with
 largest absolute value.
 Spectral properties of fermion matrices are shortly discussed in the
 next section.

\section{SPECTRAL PROPERTIES}
 As is well known from quenched studies (see for instance
 \cite{CHIRALITY}), the good chiral properties of domain wall fermions
 are realized only if the four-dimensional fermion matrix $D$ used
 for the construction of the domain-wall fermion matrix $D_F$ does not
 have very small eigenvalues.
 The hermitean four-dimensional fermion matrix
 $\tilde{D} \equiv \gamma_5 D$ should have a ``gap'' near zero in its
 spectrum.

\begin{figure}[htb]
\vspace{-1.5em}
\begin{center}
\epsfig{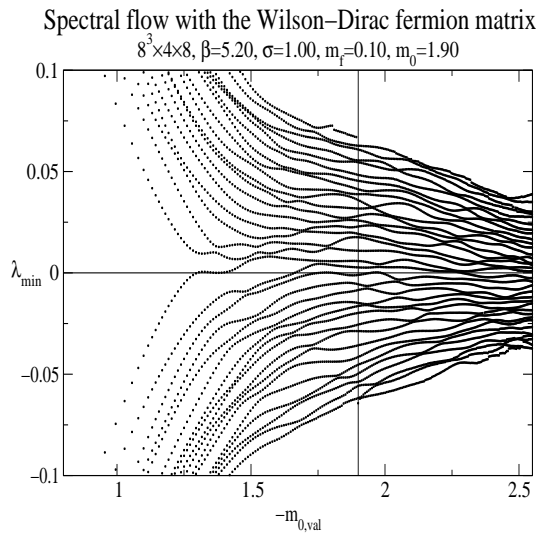}
\end{center}
\vspace{-3.5em}
\caption{\label{fig:wilsonflow}
 Eigenvalue flow for the hermitean Wilson-Dirac matrix.}
\vspace{-2.0em}
\end{figure}
\begin{figure}[htb]
\vspace{0.5em}
\begin{center}
\epsfig{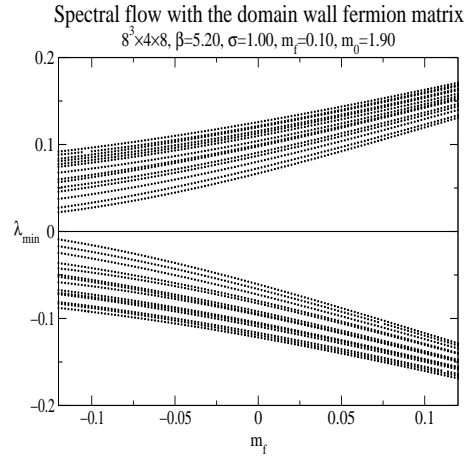}
\end{center}
\vspace{-3.5em}
\caption{\label{fig:domainflow}
 Eigenvalue flow for the hermitean domain wall fermion matrix.}
\vspace{-1.5em}
\end{figure}

 In our unquenched test runs such a gap does not appear (see figure
 \ref{fig:wilsonflow}).
 The lattice spacing is obviously still too large.
 The spectrum flow of $D_F$ on the smae gauge configuration is shown by
 figure \ref{fig:domainflow}.

 The conclusion of our first tests is that the application of the TSMB
 algorithm for numerical simulations of domain wall quarks is
 straightforward.
 A comparison of the computation speed compared to, say, Wilson quarks
 requires an analysis including the measurement of autocorrelations.
 Since the good chiral properties of domain wall fermions develop
 only sufficiently close to the continuum limit, the performance
 studies have to be carried out on large lattices.



\begin{thebibliography}{9}
%
\bibitem{KAPLAN}
D.B. Kaplan,
Phys. Lett. B288 (1992) 342.
%
\bibitem{NARANEU}
R. Narayanan, H. Neuberger,
Phys. Lett. B302 (1993) 62.
%
\bibitem{SHAMIR}
Y. Shamir,
Nucl.Phys. B406 (1993) 90.
%
\bibitem{TSMB}
I. Montvay,
Nucl. Phys. B466 (1996) 259.
%
\bibitem{POLYNOM1}
I. Montvay,
Comput. Phys. Commun. 109 (1998) 144.
%
\bibitem{POLYNOM2}
I. Montvay,
in ``Numerical challenges in lattice quantum chromodynamics'',
Wuppertal 1999, Springer 2000, p. 153; hep-lat/9911014.
%
\bibitem{SYMREV}
I. Montvay, 
Int. J. Mod. Phys. A17 (2002) 2377.
%
\bibitem{LIGHTQ}
qq+q Collaboration, F. Farchioni et al.,
to appear in Eur. Phys. J. C; hep-lat/0206008.
%
\bibitem{LIGHTB}
qq+q Collaboration, F. Farchioni et al.,
these proceedings.
%
\bibitem{DOMAINWALL}
I. Montvay, 
Phys. Lett. B537 (2002) 69.
%
\bibitem{VRANAS}
P.M. Vranas,
Phys. Rev. D57 (1998) 1415.
%
\bibitem{MULTIB}
M. L\"uscher,
Nucl. Phys. B418 (1994) 637.
%
\bibitem{NT4}
P. Chen et al.,
Phys. Rev. D64 (2001) 014503.
%
\bibitem{CHIRALITY}
CP-PACS Collaboration, A. Ali Khan et al.,
Phys. Rev. D63 (2001) 114504.
%
\end{thebibliography}
\end{document}